\documentclass[12pt,preprint]{aastex62}
\usepackage{epstopdf}
\usepackage{mathrsfs}
\usepackage{graphicx}
\usepackage{multirow}
\usepackage{amsmath}
\usepackage{threeparttable}
\usepackage{float}
\usepackage{epstopdf}
\received{}
\revised{}
\accepted{}
\submitjournal{ApJ}

\shorttitle{On the hard $\gamma$-ray spectrum of the supernova remnant G106.3+2.7}
\shortauthors{Bao \& Chen}

\begin{document}
\title{\Large{\textbf{On the hard $\gamma$-ray spectrum of the potential PeVatron supernova remnant G106.3+2.7}}}

\correspondingauthor{Yang Chen}
\email{ygchen@nju.edu.cn}

\author{Yiwei Bao}
\affil{Department of Astronomy, Nanjing University, 163 Xianlin Avenue, Nanjing 210023, China}

\author{Yang Chen}
\affil{Department of Astronomy, Nanjing University, 163 Xianlin Avenue, Nanjing 210023, China}
\affiliation{Key Laboratory of Modern Astronomy and Astrophysics, Nanjing University, Ministry of Education, Nanjing, China}




\begin{abstract}
The Tibet AS$\gamma$ experiment has measured $\gamma$-ray flux of supernova remnant G106.3+2.7 up to 100 TeV, suggesting it {being} potentially a ``PeVatron''. Challenge arises when the hadronic scenario requires a hard proton spectrum (with spectral index $\approx 1.8$), while {usual observations and numerical simulations prefer} a soft proton spectrum {(with spectral index $\geq 2$)}. In this paper, we explore an alternative scenario to explain the $\gamma$-ray spectrum of G106.3+2.7 within the current understanding of acceleration and escape processes. We consider that the cosmic ray {particles} are scattered by the turbulence driven via Bell instability. The resulting hadronic $\gamma$-ray spectrum is novel, dominating the contribution to the emission above 10\,TeV,  and can explain the bizarre broadband spectrum of G106.3+2.7 in combination with leptonic emission from the remnant.
\end{abstract}


\section{INTRODUCTION}\label{sec:intr}
Galactic cosmic rays {(CRs)} are mostly charged particles (mainly protons) with energy up to the {so called} ``knee'' {($\sim1$--$10$ PeV)}. Based on energetic arguments, supernova remnants (SNRs) are usually believed to be the accelerators of {Galactic} CRs. However, although a large number of SNRs have been detected in $\gamma$-rays, none of {them} has been confirmed to be a PeV particle accelerator, as called ``PeVatron'' \citep[see e.g.,][]{2013MNRAS.431..415B}. Discovered in the DRAO Galactic-plane survey \citep{1990A&AS...82..113J}, G106.3+2.7 is {a} cometary SNR with a tail in the southwest and a compact head (containing PSR J2229+6114) in the northeast, {at a distance of 800 pc away from Earth \citep{2001ApJ...560..236K}}. Recently, HAWC and {the Tibet} AS$\gamma$ experiments have reported the $\gamma$-ray spectrum of SNR G106.3+2.7 above 40 TeV, arguing that {the remnant} is a promising PeVatron candidate \citep{2020ApJ...896L..29A,2021NatAs...5..460T}. In particular, \citet{2021NatAs...5..460T} for the first time measured the $\gamma$-ray flux up to 100 TeV, finding that the centroid of $\gamma$-ray emissions {deviates from the pulsar at a confidence of $3.1\sigma$, and is well correlated with a molecular cloud (MC).}  {The offset of the $\gamma$-ray emission centroid from PSR J2229+6114 is measured to be} 0.44$^\circ$ ($\sim6$ pc at a distance of 800 pc).

Both leptonic and hadronic models have been proposed to give a plausible explanation to the spectral energy distribution (SED) of the SNR. In the leptonic scenario, the electrons are suggested to be transported to its current position from the pulsar wind nebula \citep[PWN,][]{2021NatAs...5..460T,2020ApJ...897L..34L}, or be accelerated by the {blast wave} directly \citep{2021NatAs...5..460T,2021Innov...200118G}; in the hadronic scenario, the protons are suggested to be accelerated by the {blast wave} in earlier ages, or re-accelerated by the PWN adiabatically {\citep{2018MNRAS.478..926O,2021NatAs...5..460T}}. {The large offset between PSR J2229+6114 and the $\gamma$-ray emission centroid indicates that the $\gamma$-rays are more likely to be mainly contributed from the particles accelerated by the {blast wave} in the {southwestern} ``tail'' region \citep{2021NatAs...5..460T}.} In order to figure out the origin of {the} $\gamma$-rays, \citet{2021Innov...200118G} separate the PWN-dominated X-ray emitting region in the northeast from the other ({the ``tail''}) part of the SNR using XMM-Newton and Suzaku observations. They found that a pure leptonic model can hardly fit the {radio, }X-ray and $\gamma$-ray spectral data of the {``tail'' region} simultaneously. Therefore, {there must be a hadronic component in the $\gamma$-ray spectrum, and} the SED can only be explained with a hadronic \citep[with a proton spectral index $\approx 1.8$,][]{2021NatAs...5..460T} or a { leptonic-hadronic hybrid} \citep[with a proton index $\approx 1.5$,][]{2021Innov...200118G} model. However, challenge arises when hard proton spectrum is required in both hadronic and { leptonic-hadronic hybrid} models, while numerical simulations show that diffusive shock acceleration can only give rise to a soft proton spectrum \citep[with spectral index $\geq 2$, see e.g.,][]{2020ApJ...905....2C}.

\citet{2021NatAs...5..460T} suggest that the very hard proton spectrum can be { formed} in very efficient acceleration (which seems to be an extreme case) and after a very slow particle diffusion. We here explore an alternative self-consistent scenario in which the $\gamma$-ray spectrum can be explained more naturally. The magnetic field in the upstream of the {blast wave} is believed to be amplified via the non-resonant Bell instability \citep[][]{2004MNRAS.353..550B,2009MNRAS.392.1591A}: protons escaping from the upstream of the shock can drive non-resonant turbulence whose scale is smaller than the Lamour radius of the escaped particles, and the relatively-low-energy particles are thus diffused by the magnetic field amplified by the non-resonant turbulence \citep[for reviews, see e.g.,][]{2013A&ARv..21...70B}. Based on the numerical simulations \citep{2004MNRAS.353..550B,2013MNRAS.431..415B}, we calculate the escape process from the first principle instead of a model with apriori phenomenological assumptions. The resulting proton spectrum is novel and can explain the hard $\gamma$-ray spectrum well {in combination with leptonic emission from the SNR}. The model is described in \S 2 and the SED of the {remnant} is fitted in \S 3, discussion is presented in \S 4 and the conclusion is drawn in \S 5.

\section{MODEL DESCRIPTION}


For simplicity, we follow the approximation that the global maximum energy {$E_{\rm max,global}$} is reached at the beginning of the Sedov phase in which $R_{\rm sh} \propto t^{2/5}$ \citep[see e.g.,][]{2012MNRAS.427...91O}. The maximum proton energy can be given by \citep{2013MNRAS.431..415B}
\begin{equation}
\label{eq:Emax}
E_{\rm esc}(t)=230\,n_{\rm e}^{1/2} \left(\frac{\eta}{0.03}\right) \left(\frac{v_{\rm sh}}{10^4 {\rm km s^{-1}}}\right)^2\left(\frac{R_{\rm sh}}{\rm pc}\right)\,{\rm TeV},
\end{equation}
where $\eta$ is the acceleration efficiency, $n_{\rm e}$ the electron number density of the interstellar medium (ISM, {we assume that {the ISM near the SNR are fully {ionized}}), ${v_{\rm sh}}$ the velocity of the shock, and ${R_{\rm sh}}$ the radius of the shock.

Following \citet{2015APh....69....1C}, we calculate the spectrum of escaped {CR} protons in the Sedov phase. In the upstream, the protons are scattered by the wave generated via escaping of {the} higher-energy protons, and therefore there is no wave upstream to scatter the protons with the highest energy $E_{\rm esc}(t)$ \citep[][]{2004MNRAS.353..550B,2013MNRAS.431..415B}. Hence, the latter escape the system quasi-ballistically at a speed $\sim c$, inducing a current {$j_{\rm CR}=n_{\rm CR}ev_{\rm sh}$} at the shock \citep[{where $n_{\rm CR}$ represents the number density of the CRs}]{2004MNRAS.353..550B}. The {differential} number of the escaped protons with energy $E_{\rm esc}(t)$ can thus be evaluated via
\begin{equation}
e\frac{{\rm d}N_{\rm esc}(E_{\rm esc})}{{\rm d}E_{\rm esc}}\,{\rm d}E_{\rm esc} = 4\pi R^2_{\rm sh}j_{\rm CR}\,{\rm d}t.
\end{equation}
{We further assume that {the} CR pressure at the shock is a fixed fraction $\xi$ of the ram pressure}
\begin{equation}
j_{\rm CR}=n_{\rm CR}ev_{\rm sh} = \frac{e\xi \rho v_{\rm sh}^3}{E_0\Psi(E_{\rm esc})},
\end{equation}
{where}
\begin{equation}
\Psi=\begin{cases}
(E_{\rm esc}/E_0)\textup{ln}(E_{\rm esc}/E_0) & \text{$\alpha$=2}\\
\frac{\alpha-1}{\alpha-2}\left(\frac{E_{\rm esc}}{E_0}\right)^{\alpha-1} \left[1-(E_0/E_{\rm esc}^{\alpha-2})\right]& \text{$\alpha>2$},
\end{cases}
\end{equation}
{$\alpha$ is the power-law index of the parent CR spectrum at the shock, and $E_0$ the minimum {energy} of protons.}

Finally, $N_{\rm esc}$ can be expressed as \citep[see][for derivation]{2015APh....69....1C}
\begin{equation}
\frac{{\rm d}N_{\rm esc}(E_{\rm esc})}{{\rm d}E_{\rm esc}}=\frac{4\pi R_{\rm sh}^2j_{\rm CR}}{e}\frac{\textup{d}t}{\textup{d}E_{\rm esc}} = \frac{4\pi \xi \rho v^2_{\rm sh}R^2_{\rm sh}}{E_0\Psi}\frac{{\rm d}R}{{\rm d}\Psi}\frac{{\rm d}\Psi}{{\rm d}E_{\rm esc}} \propto
\begin{cases}
\rho v^2_{\rm sh}R^3_{\rm sh}E^{-2}_{\rm esc}\left[\frac{1+{\rm ln}(E_{\rm esc}/E_0)}{{\rm ln}(E_{\rm esc}/E_0)^2}\right] & \text{$\alpha=2$}\\
\rho v^2_{\rm sh}R^3_{\rm sh}\left(E_{\rm esc}/E_0\right)^{-\alpha} & \text{$\alpha>2$}.
\end{cases}
\end{equation}

\section{Application to SNR G106.3+2.7}
\subsection{$\gamma$-ray spectroscopic luminosity}
{Since kinematic/physical signature of direct MC-SNR contact seems inconclusive yet \citep{LiuQC}, the MC may only be illuminated by the escaped protons. For such a scenario, the MC is sometimes approximated as a truncated cone \citep[see e.g.,][]{2012MNRAS.421..935L,2019MNRAS.490.4317C}. Here the MC is assumed to subtend a solid angle $\Omega$ at the SNR center with an inner radius $R_1$ and an outer radius $R_2$.} The diffusion coefficient near the SNR is adopted to be {$D(E)=D_{\rm 100\,TeV}(E/100\,{\rm TeV})^\delta$, where $D_{\rm 100\,TeV}$ is the diffusion coefficient for particles at 100 TeV, and $\delta$ the energy dependence index.}

The diffusion equation for {the} protons with energy $E_{\rm esc}$ escaping the shock when the shock radius was $R_{\rm sh}$ at time $T_{\rm esc}$ writes
\begin{equation}
\label{eq:dif}
\frac{\partial}{\partial t}f(E_{\rm esc},r,t)=\\ \frac{D(E_{\rm esc})}{r^{2}}\frac{\partial}{\partial r}\left[r^{2}\frac{\partial}{\partial r}f(E_{\rm esc},r,t)\right]+Q,
\end{equation}
where
\begin{equation}
\begin{aligned}
Q & =\frac{1}{4\pi R_{\rm sh}^2}\frac{{\rm d}N_{\rm esc}(E_{\rm esc})}{{\rm d}E_{\rm esc}}\delta(r-R_{\rm sh})\delta(t-T_{\rm esc}) \\ &=\frac{C_{\rm Q}\rho v^2_{\rm sh}R^3_{\rm sh}(E_{\rm esc}/E_0)^{-2}}{4\pi R^2_{\rm sh}}\delta(r-R_{\rm sh})\delta(t-T_{\rm esc}) \times
\begin{cases}
\left[\frac{1+{\rm ln}(E_{\rm esc}/E_0)}{{\rm ln}(E_{\rm esc}/E_0)^2}\right] & \text{$\alpha=2$}\\
(\alpha-2)\left(E_{\rm esc}/E_0\right)^{-\alpha+2} & \text{$\alpha>2$}
\end{cases}
\end{aligned}
\end{equation}
is the injection term, $f$ the proton distribution function, and} $C_{\rm Q}$ a {free parameter which is proportional to the total proton energy}.


{As is shown in \autoref{app:A}, the} solution of \autoref{eq:dif} is \citep[see also][]{1995PhRvD..52.3265A,2019MNRAS.490.4317C}

\begin{equation}\label{eq:sol}
\begin{aligned}
f(E_{\rm esc},r,t) = \frac{{{\rm d}N_{\rm esc}(E_{\rm esc})}/{{\rm d}E_{\rm esc}}}{4 r \pi^{3/2} R_{\rm sh} R_{\rm dif}} \left\{{\rm exp}\left[-\left(\frac{r-R_{\rm sh}}{R_{\rm dif}}\right)^2\right]-{\rm exp}\left[-\left(\frac{r+R_{\rm sh}}{R_{\rm dif}}\right)^2\right]\right\},
\end{aligned}
\end{equation}
{where $R_{\rm dif}=2\sqrt{D(E_{\rm esc})(t-T_{\rm esc})}$ is the diffusion length scale.}

At time $T_{\rm age}$, the differential number of protons with energy $E_{\rm esc}$ lying inside the conic MC shell can thus be calculated to be
\begin{equation}
N_{\rm CR,MC}=\int_{R_1}^{R_2} f(E,r,t) \Omega r^2\,{\rm d}r.
\end{equation}


{ Finally, the hadronic $\gamma$-ray spectroscopic luminosity {$\Phi_{\gamma}(E_{\gamma})$} is evaluated via the {cross section} presented in \citet{2014PhRvD..90l3014K}

\begin{equation}
\Phi_{\gamma}(E_{\gamma}) = cn_{\rm MC} \int \frac{{\rm d}\sigma}{{\rm d}E_{\gamma}}(E,E_{\gamma}) N_{\rm CR,MC}(E) {\rm d}E,
\end{equation}
where ${\rm d}\sigma/{\rm d}E_{\gamma}$ is the $\gamma$-ray differential cross section.
}

In addition to the hadronic component, we add a leptonic component contributed by the electrons accelerated by the {blast wave}. We approximate the spectrum of electrons in the SNR to be a broken power-law
\begin{equation}
\frac{{\rm d}N_{\rm e}}{dE}\propto
\begin{cases}
E^{-\alpha_1}& {E\leq E_{\rm b}}\\
E^{-\alpha_2}& {E_{\rm b}<E<\min\left(E_{\rm max,global},E_{\rm loss}\right)},
\end{cases}
\end{equation}
{where ${\rm d}N_{\rm e}/{\rm d}E$ the differential number of electrons, and $E_{\rm loss}$ the maximum energy determined by energy loss;
$\alpha_1$ and $\alpha_2$ are the power-law indices in the low energy and high energy band, respectively;
{ $E_{\rm b}$, the break energy of the electron spectrum, can be constrained well \citep[$\sim9$ TeV, assuming a magnetic field of 6 $\mu$G,][]{2021Innov...200118G} by the radio \citep{2000AJ....120.3218P} and X-ray {\citep{2021Innov...200118G,2021ApJ...912..133F}} emissions, which are both dominated by the SNR. The electron spectrum above the break is quite soft, and the leptonic $\gamma$-ray emissions are dominated over by the hadronic emissions above 10 TeV. Hence, the total $\gamma$-ray spectrum is insensitive to the electron spectrum above $E_{\rm b}$.}

We fit the SED as is plotted in \autoref{fig:SED}, with the parameters listed in \autoref{tab:par}.} {We find that the parameters are insensitive to X-ray flux, since the different sets of X-ray flux data from \citet{2021Innov...200118G} and \citet{2021ApJ...912..133F} can be fitted with very similar parameters.}

{In \autoref{fig:pMC}, we plot the proton spectrum insides the MC. As can been seen, the protons with energy $<{75}$ TeV are still trapped by the turbulence driven by the escaping protons via Bell instability, absent in the MC. Hence, only protons with {energies} ranging from ${75}$ TeV to 280 TeV can reach the MC {within the relatively short lifetime of the SNR} and contribute to a novel hadronic $\gamma$-ray spectrum. {In \autoref{fig:variation}, we explore the dependence of hadronic $\gamma$-ray spectrum on $\alpha$ and $\delta$ and find that the choice of $\alpha$ (in the range 2.0--2.6) or $\delta$ (in the range of 0--1) does not impact the results significantly.
The reason is that the protons which can hit the MC within the relatively short lifetime of the SNR at a narrow energy range (75--280 TeV) is almost mono-energetic (as shown in \autoref{fig:pMC}). Hence, the energy dependence index $\delta$ and the power-law index $\alpha$ can hardly affect the proton spectrum in the MC. \citet{2019MNRAS.490.4317C} have also proposed a similar phenomenological escaping scenario which can lead to a hard hadronic $\gamma$-ray spectrum in middle-aged SNRs, while in our case, we model a younger SNR from the first principle.}

Since the hadronic $\gamma$-ray luminosity $\Phi_{\gamma} \propto C_{\rm Q} n_{\rm MC}\Omega$ (where $n_{\rm MC}$ is the number density of {the gas in} the MC, a free parameter), we only list $C_{\rm Q} n_{\rm MC}(\Omega/4\pi)$ as a single parameter in \autoref{tab:par}. {Once $C_{\rm Q} n_{\rm MC}\Omega$ is obtained from data fitting, the total proton energy $E_{\rm p,tot}$ (including that in the GeV protons trapped near the shock) can be calculated via}
\begin{equation}
E_{\rm p,tot} \approx \int_{E_0}^{E_{\rm max,global}}  E\,{\rm d}E \cdot C_{\rm Q}\rho v^2_{\rm sh}(T_{\rm age}) R^3_{\rm sh}(T_{\rm age}) \left(\frac{E}{E_0}\right)^{-2} \times
\begin{cases}
\left[\frac{1+{\rm ln}(E_{\rm esc}/E_0)}{{\rm ln}(E_{\rm esc}/E_0)^2}\right] & \text{$\alpha=2$}\\
(\alpha-2)\left(E_{\rm esc}/E_0\right)^{-\alpha+2} & \text{$\alpha>2$}.
\end{cases}
\end{equation}
{If we adopt $n_{\rm MC}=100$ cm$^{-3}$, $E_{\rm p,tot}$ will be $1.6 \left(4\pi/\Omega\right) \times 10^{49}$ erg, which is quite reasonable}.

{ As shown in \autoref{fig:SED}, the leptonic component accounts for the radio \citep{2000AJ....120.3218P} and X-ray (Ge et al. 2021 for Model A; Fujita et al. 2021 for Model B) emission and dominates the $\gamma$-ray flux below 500\,GeV. Meanwhile, the hadronic $\gamma$-rays (the yellow {lines}) have a very hard spectrum (d$N_{\gamma}/{\rm d}E_{\gamma} \propto E_{\gamma}^{-1}$) below 500 GeV, { which stems from the proton cutoff at 75 TeV and the low-energy tail of the pp interaction cross section}.} The spectrum above 10\,TeV is dominated by the hadronic component, and the whole $\gamma$-ray spectrum can be explained by the hybrid model naturally.

\subsection{Age of SNR G106.3+2.7}

There remains some uncertainty about the age of {SNR G106.3+2.7}: although the characteristic age of PSR J2229+6114 ($\approx 10^4$ yr) is usually adopted to be the age of the {remnant}, sometimes the {remnant} is suggested to be very young \citep[$\sim 1$ kyr, see e.g.,][]{2020ApJ...896L..29A}. With an adiabatic expansion ($R_{\rm sh}\approx 1.2 (E_{\rm SN}/\rho_{\rm ISM})^{1/5} t^{2/5}$, where $\rho_{\rm ISM}$ is the density of the ISM), {the estimated $E_{\rm SN} \approx 7\times 10^{49}$ erg \citep{2001ApJ...560..236K} is unusually low} if the age of {remnant} is $\approx 10$ kyr. However, as is estimated in \citet{2015APh....69....1C}, \autoref{eq:Emax} indicates that the global maximum energy $E_{\rm max,global} \propto E_{\rm SN}$, and {SNRs} with {a} low $E_{\rm SN}$ can hardly accelerate CRs to $\sim 10^2$ TeV. In order to reconcile the {dilemmas}, in this paper we consider a spherically symmetric scenario in which $E_{\rm SN}=10^{51}$ erg (the canonical value) and $T_{\rm age}=1000$ yr. Such {a remnant} age is plausible for a {hosted} pulsar with a characteristic age of $\sim 10^4$ yr.
Assuming a canonical braking index {n:=$\nu \ddot\nu/\dot\nu^2=3$ (where $\nu$ is the spin frequency of the pulsar)}, the real age of the pulsar writes $T_{\rm age}=\tau_{\rm c}[1-(P_0/P_{\rm now})^2]$, where $\tau_c$ is the characteristic age of the pulsar, $P_0$ the {initial} spin period of the pulsar, and $P_{\rm now}$ the spin period of the pulsar at present.
For PSR J2229+6114, $P_{\rm now}=50$ ms \citep{2001ApJ...552L.125H}, and $T_{\rm age}$ can be as small as $\sim 1$ kyr if $P_0$ is {appropriately} close to $P_{\rm now}$.
{Such value of $P_0$ is allowed,} since in pulsar evolution models $P_0$ is usually suggested to be in a wide range from $\sim 4$ ms to $\sim 400$ ms \citep[see e.g.,][]{2002ApJ...568..289A,2006ApJ...643..332F,2010MNRAS.401.2675P}. Instances in which $P_0\approx P_{\rm now}$ and $T_{\rm age} \ll \tau_c$ can also be found in the catalog: (a) CCO 1E 1207.4$-$5209 inside SNR G296.5+10.0 has $P_0 \approx P_{\rm now}=424$ ms, leading to $\tau_{\rm c}>27$ Myr, which exceeds the age of the SNR by 3 orders of magnitude \citep{2007ApJ...664L..35G}, and (b) PSR J1852$-$0040 is suggested to have $P_0 \approx P_{\rm now} \approx 10^2$ ms, giving rise to $\tau_{\rm c} \approx 2\times 10^8$ yr, 4 orders of magnitude larger than $T_{\rm age} \sim 5\times 10^3$ yr measured from the observation {of the} associated SNR \citep{2005ApJ...627..390G}. On the other hand, as can be seen from \autoref{eq:Emax}, $E_{\rm esc}$ is determined by the $R_{\rm sh}$ and $v_{\rm sh}$ which can be constrained by observations directly, irrespective of $T_{\rm age}$. \citet{2021Innov...200118G} showed that $v_{\rm sh}$ is at least 3000 km s$^{-1}$ in the {southwestern tail} region based on {the} non-thermal X-ray spectrum up to 7 keV without a clear spectral cutoff, while $R_{\rm sh}\approx 6$ pc can be estimated via radio observations \citep[see e.g.,][]{2021NatAs...5..460T}. Although $T_{\rm age}$ does affect $R_{\rm dif}$, its impacts can be compensated by the free parameter of {$D_{\rm 100\ TeV}$}. Hence, for simplicity we only consider a spherically symmetric {remnant} evolving {in uniform medium} following the well-known \citet{1999ApJS..120..299T} model\footnote{See \citet{2017AJ....153..239L} for a fast python calculator on SNR evolution.
}.

\subsection{Other issues}

{A sharp cutoff in the novel proton spectrum is an interesting characteristic in our particle escaping scenario with a spherically symmetric morphology applied, which can well fit the {Tibet} AS$\gamma$ data in this SNR.}
Meanwhile, {instead of a sharp cutoff,} smoother break can be predicted if {an} asymmetric morphology is {considered}. {In that case,} the shock radius $R_{\rm sh}$ and velocity $v_{\rm sh}$ both vary in different directions, and thus $E_{\rm esc} \propto v_{\rm sh}^2R_{\rm sh}$ also varies in different direction accordingly{, and thus} the spatial variation of $E_{\rm esc}$ {may turn} the sharp cutoff into a softer break.
Since the data available at present can not distinguish the asymmetric model, here we adopt a symmetric model {to explain the $\gamma$-ray spectrum in the zeroth order}. Further observation carried {out} by LHASSO may help to distinguish these models.} During the lifetime of {SNRs}, the shock may break and the CRs can thus escape with continuous power-law spectra, which is the case as previously adopted \citep[see e.g.,][]{2010MNRAS.409L..35L,2011MNRAS.410.1577O}. In the SNR catalog\footnote{http://snrcat.physics.umanitoba.ca/SNRtable.php} \citep{2012AdSpR..49.1313F}, there are only a few SNRs with $\gamma$-ray emissions in $\la 10$ TeV {which} are confirmed to be of hadronic origin, and our novel hadronic spectrum may be expected to be found in more SNRs with the help of {LHASSO \citep[see e.g.,][]{2021ChPhC..45b5002A}, {Tibet} AS$\gamma$, HAWC, CTA \citep[see e.g.,][]{2021JCAP...02..048A} and ASTRI Mini-Array \citep[see e.g.,][]{2020JHEAp..26...83P} experiments} in future.

\section{summary}
{Although the standard theory of non-linear diffusive shock acceleration can predict hard proton spectrum with $\alpha<2$ \citep{2020ApJ...905....2C}, it is noted that observations and recent numerical simulations seem to prefer a soft hard proton spectrum \citep{2020ApJ...905....2C}.} In this paper, we explore an alternative plausible scenario to explain the {bizarrely hard} $\gamma$-ray spectrum of SNR G106.3+2.7 within the current acceleration theory which {predicts} soft ($\alpha \geq 2$) proton spectra.
Apart from the common leptonic component which can be constrained by radio and X-ray observations, we {invoke} a novel hadronic component. {This} component arises from the escape scenario proposed by \citet{2004MNRAS.353..550B} and \citet{2015APh....69....1C}, in which the CRs at the shock are diffused by the turbulence {that is} generated by {the escaping CR particles with higher energies}. Consequently, at {a given} time, only CRs with {the highest} energy can escape upstream, while CRs with lower {energies} are confined. {Therefore, only protons with energies {between} 75--280 TeV can reach the MCs at {an age 1 kyr}. {The cutoff of the proton spectrum at 75 TeV then gives rise to a very hard hadronic $\gamma$-ray spectrum below 10 TeV because of the low-energy tail of the pp cross section.} The hadronic $\gamma$-ray spectrum, which dominates above {10\,TeV}, together with the leptonic component, can explain the bizarrely hard $\gamma$-ray spectrum of SNR G106.3+2.7 well.

In our model treatment, the $\gamma$-ray spectrum is calculated in zeroth order, assuming that a spherically symmetric SNR expands in a uniform medium.
There are seven free parameters in the hadronic component, namely, $n_{\rm ISM}$, $D_{\rm 100 TeV}$, $R_1$, $R_2$, $M_{\rm ej}$, $\eta$, and $n_{\rm MC} C_{\rm Q}(\Omega/4\pi)$.
Neither of parameters $\alpha$ and $\delta$ has strong correlation to the hadronic $\gamma$-ray spectrum, because the protons which can hit the MC within the relatively short lifetime of the SNR in the narrow energy range (75--280\,TeV) are approximately mono-energetic.



\appendix
\section{analytical solution of diffusion equation}\label{app:A}
{In order to solve \autoref{eq:dif} analytically, we define a new function $F := rf(E_{\rm esc},r,t)$, and then \autoref{eq:dif} reads}
\begin{equation}
\label{eq:dif2}
\frac{\partial}{\partial t} F=D(E_{\rm esc})\frac{\partial^2 F}{\partial r^2}+Q(E_{\rm esc},r,t)r,
\end{equation}
{and the boundary condition is $\partial f/\partial r =0$ and $F|_{\rm r=0}=0$. Defining the differential operator}
\begin{equation*}
\mathcal{L} := \frac{\partial }{\partial t}-D\frac{\partial^2}{\partial r^2},
\end{equation*}
{\autoref{eq:dif2} can be further written as $\mathcal{L}F=Qr$. We firstly discuss a simpler equation $\mathcal{L}G(r,\zeta)=\delta(r-\zeta)$ with the boundary condition $G|_{\rm r=0}=0$. The solution is apparently }
\begin{equation}\label{eq:Gdelta}
G(r,\zeta)=\frac{1}{\sqrt{\pi}R_{\rm dif}}\left\{\exp\left[-\left(\frac{r-\zeta}{R_{\rm dif}}\right)^2\right]-\exp\left[-\left(\frac{r+\zeta}{R_{\rm dif}}\right)^2\right]\right\}.
\end{equation}
{Then we multiply \autoref{eq:Gdelta} by $h(\zeta)=Q(\zeta)\zeta$ on both sides, integrate over $\zeta$, and exchange $\mathcal{L}$ and $\int$, i.e.,}
\begin{equation}
    \mathcal{L}\left( \int^\infty_0 G(r,\zeta)Q(\zeta)\zeta\ {\rm d}\zeta \right) = \int^\infty_0 \mathcal{L}G(r,\zeta)Q(\zeta)\zeta\ {\rm d}\zeta =   \int^\infty_0 \delta(r-\zeta)Q(\zeta)\zeta\ {\rm d}\zeta=Qr=\mathcal{L}F.
\end{equation}
{Hence, $\int^\infty_0 G(r,\zeta)Q(\zeta)\zeta\ {\rm d}\zeta=F$, and we thus have}
\begin{equation}
\begin{aligned}
f &=\frac{1}{r}\int_{0}^{\infty} Q(\zeta) \zeta G(E_{\rm esc},t,T_{\rm esc};r,\zeta) {\rm d}\zeta\\ &= \frac{{{\rm d}N_{\rm esc}(E_{\rm esc})}/{{\rm d}E_{\rm esc}}}{4 r \pi^{3/2} R_{\rm sh} R_{\rm dif}} \left\{{\rm exp}\left[-\left(\frac{r-R_{\rm sh}}{R_{\rm dif}}\right)^2\right]-{\rm exp}\left[-\left(\frac{r+R_{\rm sh}}{R_{\rm dif}}\right)^2\right]\right\}
\end{aligned}
\end{equation}

\begin{center}
\begin{figure}
\includegraphics[scale=0.7]{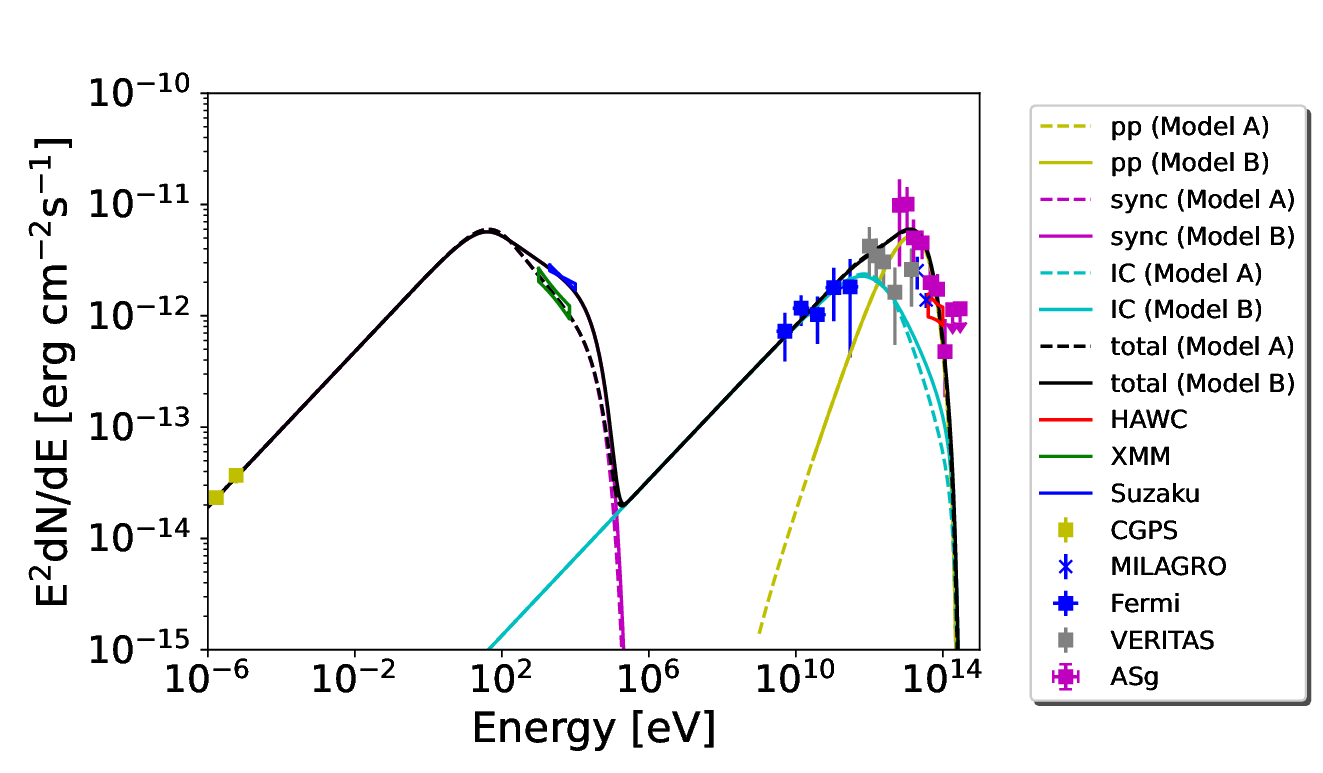}
	\caption{{SED of the emission from SNR G106.3+2.7.} The CGPS data are taken from \citet{2000AJ....120.3218P}, XMM-Newton data from {\citet[][for Model A]{2021Innov...200118G}}, {Suzaku data from \citet[][for Model B; see also \autoref{tab:par}]{2021ApJ...912..133F}}, Fermi-LAT data from \citet{2019ApJ...885..162X}, VERITAS data from \citet{2020ApJ...896L..29A}, HAWC data from \citet{2020ApJ...896L..29A}, and {Tibet} AS$\gamma$ data from \citet{2021NatAs...5..460T}. The hadronic $\gamma$-ray spectrum below 500 GeV is contributed by {75}--280 TeV protons via a small cross section.}
\label{fig:SED}
\end{figure}
\end{center}

\begin{center}
\begin{figure}
\includegraphics[scale=0.8]{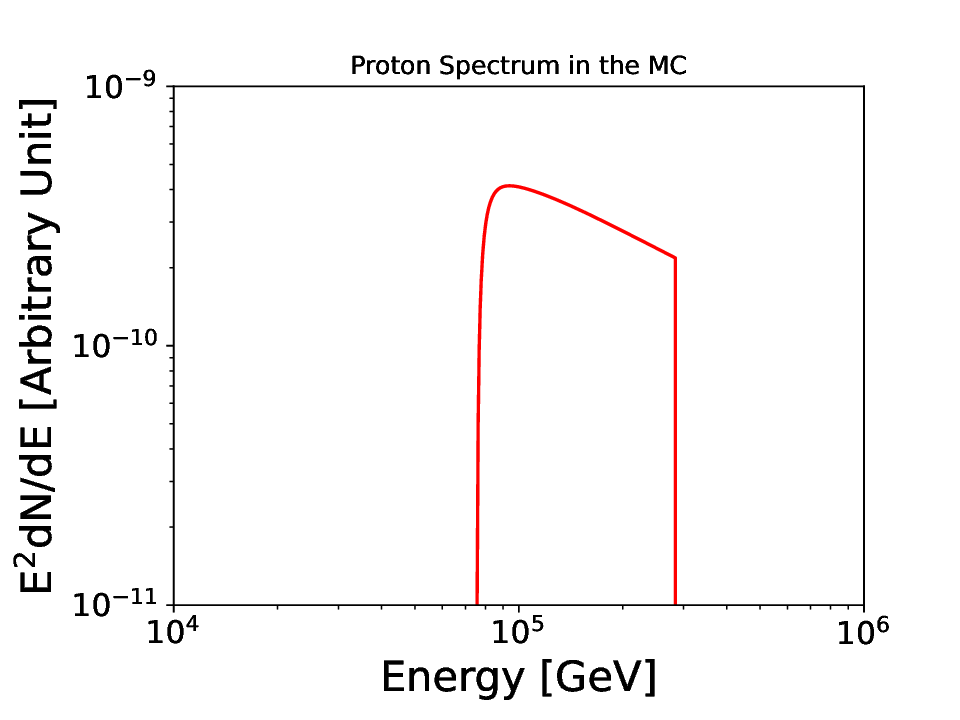}
	\caption{Proton spectrum in MC.}
\label{fig:pMC}
\end{figure}
\end{center}

\begin{center}
\begin{figure}
\includegraphics[scale=0.5]{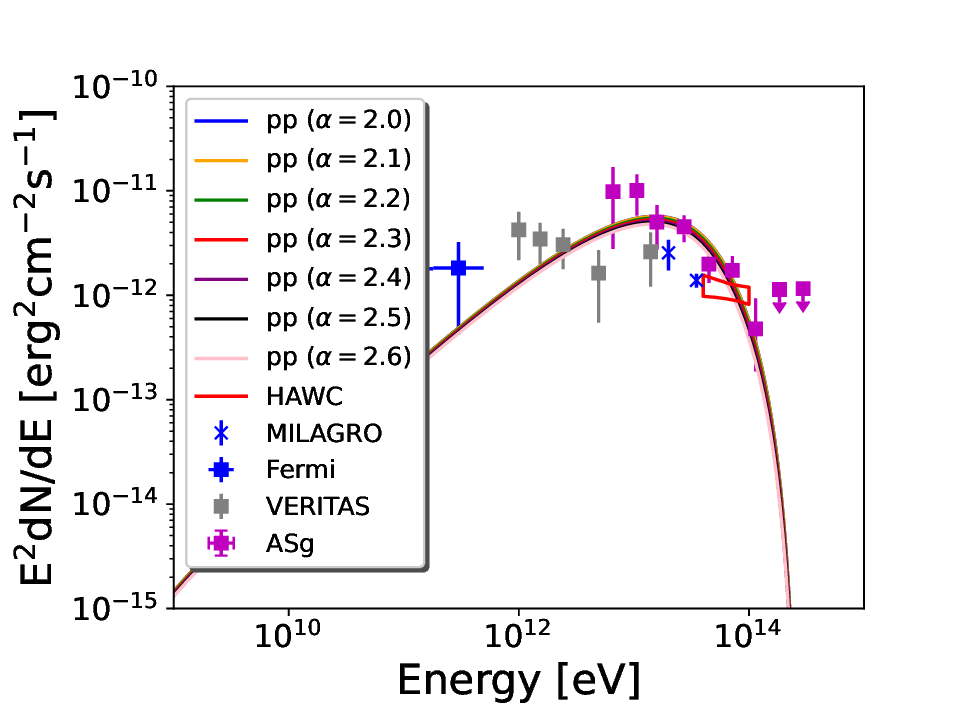}
\includegraphics[scale=0.5]{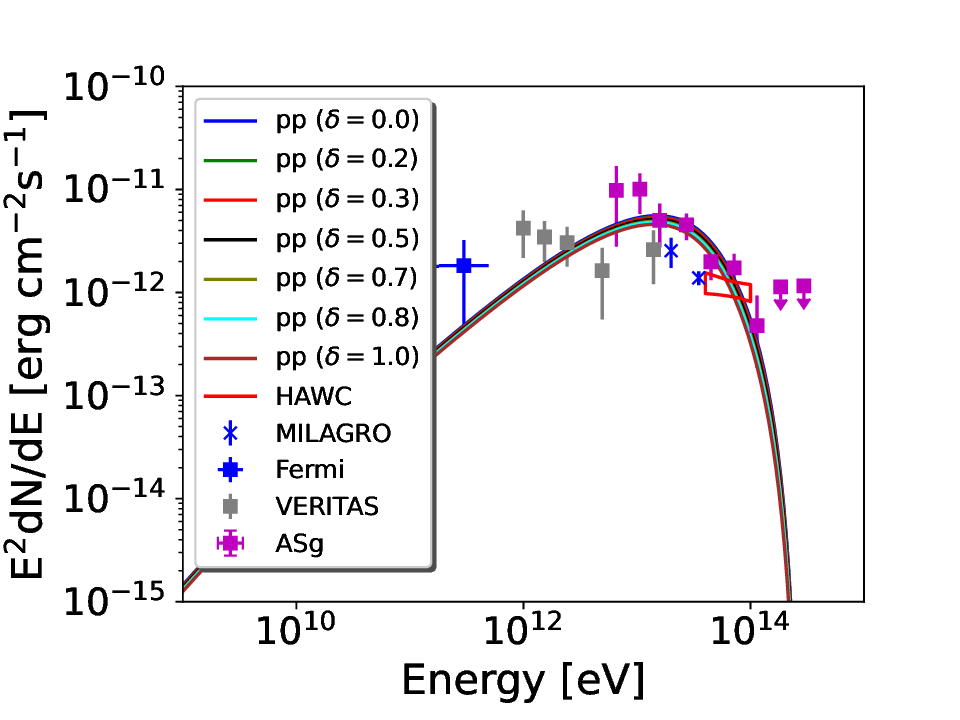}
	\caption{Dependence of the hadronic $\gamma$-ray spectrum on the $\alpha$ (left panel, with $\delta=1/3$) and $\delta$ (right panel, with $\alpha=2.3$).}
\label{fig:variation}
\end{figure}
\end{center}

\begin{center}
\begin{deluxetable}{ccc}
\tabletypesize{\footnotesize}
\tablecaption{Fitting Parameters\label{tab:par}}
\tablewidth{0pt}
\tablehead{
\\
Parameter     & Model A& Model B\\
}
\startdata
$T_{\rm age}$ (yr)                       	& 1000 & ...\\
$R_{\rm now}$ (pc)							& 5.5 & ...\\
$E_{\rm max,global}$ (TeV)					& $2.8\times10^2$  & ...\\
$E_0$ (GeV)									& 1.0  & ...\\
$\alpha$ 									& 2.3 & ...\\
$D_{\rm 100\,TeV}$ (cm$^2$ s$^{-1}$) & $6.6\times 10^{27}$ \\
$\delta$ 									& 1/3 & ... \\
$R_1$ (pc)									& 7    & ...\\
$R_2$ (pc)									& 10   & ...\\
$E_{\rm SN}$ (erg)							& $1.0 \times 10^{51}$ & ...\\
$d$ (pc)                       				& 800&...\\
$M_{\rm ej}$ ($M_\odot$)					& 1.0 &...\\
$n_{\rm e}$  (cm$^{-3}$)	            	& 0.6&...\\
$\eta$			 							& 0.04&...\\
$n_{\rm MC} C_{\rm Q}\left(\Omega/4\pi \right)$ (erg$^{-2}$ cm$^{-3}$)&$320$ & ...\\
\hline
$E_{\rm b}$ (TeV)							& 12 & 10 \\
$\alpha_1$ 									& 2.3 & ...\\
$\alpha_2$ 									& 3.8 & 3.45\\
$B$ ($\mu$G)								& 4.6 & ...\\
$W_{\rm e}$	(erg)							& $6.8\times 10^{47}$& ...\\
\hline
$T_{\rm CMB}$ (K)			             & 2.73 & ... 			\\
$u_{\rm CMB}$ (eV cm$^{-3}$)			& 0.25 	& ...		\\
$T_{\rm FIR}$ (K)					& 25		& ...		\\
$u_{\rm FIR}$ (eV cm$^{-3}$)			& 0.2 	& ...			\\
$T_{\rm OPT}$ (K)					& 3000		& ...		\\
$u_{\rm OPT}$ (eV cm$^{-3}$)			& 0.3 	& ...			\\
\enddata
\tablenotetext{*}{ $d$ is the distance to the SNR, $R_{\rm now}$ the radius of the SNR at present, $E_{\rm SN}$ the explosion energy of the SNR, $M_{\rm ej}$ the ejecta mass, $B$ the magnetic field strength, and $W_{\rm e}$ the total electron energy. $T_{\rm CMB}$, $T_{\rm FIR}$, and $T_{\rm NIR}$ are the temperatures of the CMB, far infrared, and near infrared photons, respectively; $u_{\rm CMB}$, $u_{\rm FIR}$, and $u_{\rm NIR}$ are energy densities of the CMB, far infrared, and near infrared photons, respectively.}
\end{deluxetable}
\end{center}

\acknowledgments
We are in debt to Pasquale Blasi for fruitful discussion, {and to an anonymous referee, Yutaka Fujita}, Ruo-Yu Liu, Qiang Yuan and Xiangdong Li for helpful comments. This work is supported by the National Key R\&D Program of China under grants 2017YFA0402600, NSFC under grants 11773014, 11633007, 11851305 and {U1931204}.

\end{document}